\documentclass[structabstract]{aa}  
\usepackage{graphicx}
\usepackage{txfonts}
\begin{document}
   \title{Fluorine abundances in dwarf stars of the solar neighbourhood   \thanks{Based on VLT/CRIRES observations collected at the European Southern Observatory, proposal 079.D-0450.}}

   \author{A. Recio-Blanco\inst{1}
          \and
          P. de Laverny\inst{1}
          \and
          C. Worley\inst{1}
          \and
          N. C. Santos\inst{2,3}
          \and
          C. Melo\inst{4}
          \and
          G. Israelian\inst{5}
          }

   \institute{Laboratoire Lagrange, UMR7293, Universit\'e de Nice Sophia Antipolis, CNRS, Observatoire de la C\^ote d'Azur, BP 4229, 06304 Nice, France\\
              \email{arecio@oca.eu} 
         \and
             Centro de Astrofisica da Universidade do Porto, Rua das Estrelas, 4150-762 Porto, Portugal
         \and
             Departamento de F\'isica e Astronomia, Faculdade de Ci\^encias, Universidade do Porto, Portugal 
         \and
             European Organization for Astronomical Research in the Southern Hemisphere; Alonso de C\'ordova 3107, Casilla 19001 - Santiago 19, Chile 
         \and
             Instituto de Astrof\' isica de Canarias, 38200 La Laguna, Tenerife, Spain
             }
 
   \date{Received 2011; accepted 2011}

 
  \abstract
   {}
   {In spite of many observational efforts aiming to characterize 
the chemical evolution of our Galaxy, not much is known about the origin of 
fluorine (F). Models suggest that the F found in the Galaxy might have been 
produced mainly in  three different ways, namely, Type~II Supernovae, asymptotic
giant branch nucleosynthesis, or in the core of Wolf-Rayet stars. 
Only a few observational measurements of F abundances are available 
in the literature and mostly for 
objects whose characteristics might hamper an accurate determination of fluorine
abundance (e.g.,complex mixing and nucleosynthesis processes, external/internal contamination). }
  {We derive the F abundances for a set of nine cool main-sequence dwarfs in the solar
neighbourhood, based on an unblended line of the HF molecule at 2.3 microns. In addition, 
we study the s-process elements of five of these stars.}
   {We acquire data using the high-resolution IR-spectrograph 
CRIRES and gather FEROS data from the European Southern Observatory archive. The classical method
of spectral synthesis in local thermodynamic equilibrium has been used for the abundance
analysis.}
   {Several of the analysed stars seem to be slightly fluorine enhanced with respect to the Sun, although no
correlation is found between the F abundance and the iron content.  In addition,
the most fluorine enriched stars are also yttrium and zirconium enriched, which suggests that AGB fluorine nucleosynthesis
is the dominant source of fluorine production for the observed stars. Nevertheless, the correlation between [F/Fe] 
and the s-elements is rather weak and possibly masked by the uncertainties in the F abundance measurements.
Finally, we compare our derived F abundances to previous measurements of alpha-element and iron-peak
element abundances. Type II core collapse Supernovae do not appear to be
the main site of F production for our targets, as no correlation 
seems to exist between the [F/Fe] and the [$\alpha$/Fe] ratios.
}
   {}


   \maketitle
%

\section{Introduction}
Fluorine is currently believed to come from three main different astrophysical sources. First, it 
is thought to be produced in type II supernovae through neutrino spallation of one proton of $^{20}$Ne, 
following the core-collapse phase of a massive star (e.g. \cite{Woosley}). 
Second, low-mass (2-4~M$_{\odot}$) asymptotic giant branch (AGB) stars are also believed to be producers of F during 
He-burning pulses. Several lines of evidence confirm that the nucleosynthesis of F occurs in AGB stars:
correlations  of F enhancements with other products of AGB nucleosynthesis such as the C/O ratio and 
the s-element abundances
(\cite{Jorissen}, \cite{Abia2010}); the observation of an enhancement in the fluorine content of 
planetary nebulae (\cite{Liu}, \cite{Otsuka}) or post-AGB stars (\cite{Werner}); and the detection of 
the AlF molecule in 
the circumstellar envelope of the AGB star IRC+10216 (\cite{Cernicharo}).  Moreover, there is evidence 
that fluorine is produced in the cores of Wolf-Rayet (WR) stars at the begining of their He-burning phase 
(\cite{Meynet}, \cite{Stancliffe}), and is spread into 
the interstellar medium by the strong mass-loss rate of these objects (with relatively high metal 
content). As for the AGB stars, $^{19}$F in this case is produced from $^{14}$N nucleii 
(e.g. \cite{ZL}). Nevertheless, Palacios et al. (2005), who included rotation in their models,  
showed that the WR contribution to Galactic fluorine remains quite open. 
In addition to the above-mentioned sources of F, a fourth possible contributor was
proposed by \cite{Longland}, who suggested that
F could be synthesised in the merger of a helium white dwarf and a carbon-oxygen white dwarf.
   
Which of the above-mentioned sources is the most important one is currently under debate. In 
spite of the many uncertainties regarding the origin of F in the Galaxy, there have not been many 
observational studies of this element. A pioneering study was carried out in 1992 (\cite{Jorissen}), where 
F abundances were derived from HF lines for a set of Galactic K 
and M giants, as well as a large number of s-process enriched Galactic AGB stars. All the targets 
had metallicities near solar. Since then, other works have studied the nucleosynthesis of F in AGB stars. 
Abia et al. (2009, 2010 and 2011) revised the F enhancements of the C-rich objects
in the \cite{Jorissen} sample and determined for the first time F abundances in six extragalactic
carbon stars. Their measurements are in closer agreement with the most recent theoretical nucleosynthesis
predicions (\cite{Cristallo}), although they have a different dependence on the stellar metallicity .
However, \cite{Sara} measured the F abundance of ten Galactic extrinsic carbon-rich, low metallicity stars
concluding also that predictions from nucleosynthetic models for low-mass, low-metallicity AGB stars
account for the derived F abundances, while the upper limits on the F content derived for most of the
stars are lower than the predicted values. 

The fluorine abundance patterns in the low metallicity regime were also explored by
\cite{Cunha2003}, who derived F abundances for a sample of giants in the Large Magellanic Cloud 
and in the globular cluster $\omega$ Cen. 
They found that the abundance ratio of F/O declines as the oxygen abundance decreases.
They therefore suggested that the observed low values of F/O  exclude AGB synthesis as the
dominant source of fluorine in their targeted stellar populations. In particular,
the targeted stars in $\omega$ Cen did not have enhanced F abundances, despite their large s-process 
abundance values. In addition, they claimed a
qualitative agreement with what is expected if $^{19}$F is produced 
via H- and He-burning sequences in very massive stars (fluorine being 
ejected in high mass-loss rate Wolf-Rayet winds). Finally, they found a quantitative agreement 
between the Galactic and Large Magellanic Cloud results with the predictions of models in which $^{19}$F is produced 
from neutrino nucleosynthesis during core collapse in supernovae of Type~II.

The same authors studied fluorine abundances in three 
low-mass young stars in the Orion Nebula Cluster (\cite{Cunha2005}), and in red 
giants of the southern globular cluster M4 (\cite{Smith05}). 
The later work revealed that the abundance of fluorine is found to vary by more 
than a factor of six among cluster stars. The $^{19}$F variations are correlated 
with  already established oxygen variations, and anticorrelated with the sodium 
and aluminium variations, suggesting a link with the abundance anomalies
caused by the pollution from previous generations of stars. In addition, $^{19}$F is found to decrease in the M4 stars, 
as the signature of H burning appears. Finally, \cite{Cunha2008} derived F abundances
in a sample of six bulge red giants. The lack of an s-process enhancement in the
most fluorine-rich bulge giant in this study lead the authors again to conclude that WR stars
contribute more than AGB stars to F production in the bulge and even more so than they 
do in the disc.


The afore mentioned observational results combined with theoretical models currently indicate that all three 
suggested sources of fluorine contributed to the observed abundances of this element in 
different epochs of the evolution of the Galaxy (\cite{Ren04}, \cite{Kova11}). However, the
relative contributions of the three F productors remains disputed. 
The targets observed so far are probably not the most appropriate one 
for deriving accurate F abundances. First, most of them are giant stars, where mixing may have 
occurred, changing the surface elemental ratios. In addition, some of these are globular 
cluster giants, where some important external pollution (from previous generations of stars) 
may have taken place (e.g. \cite{G00}, \cite{Smith05})
or very young cool dwarfs, which likely have infra-red disk emission.

In other words, an improvement in our knowledge of the possible fluorine astrophysical sources
is necessary to solve long-standing discrepancies between Galactic chemical-model predictions
and observed F abundances. Clues to the site of fluorine synthesis lie in the run of the fluorine 
abundance with metallicity in the Galaxy and in different stellar populations. 
We present here the fluorine abundances derived for nine field main-sequence dwarf stars in the 
solar neighbourhood, belonging to the samples studied by \cite{Santos04} and 2005 and
\cite{Sousa06}. In addition, s-process elements abundances were also determined for six of
the target stars using FEROS spectra taken from the European Southern Observatory (ESO)
archive. Section 2 describes the observations and the data analysis. The
derived fluorine and s-element abundances are presented in Section 3 and compared to the abundances of
several other elements already available for those stars from homogeneous analysis in the literature. 
Finally, conclusions are discussed in Section 4.


\section{Observations and analysis}

We used the high-resolution IR-spectrograph CRIRES (\cite{Crires})
mounted on the 8.2m Antu telescope of ESO's Very Large Telescope on Cerro Paranal
observatory to observe nine cool main-sequence dwarfs of the solar
neighbourhood.  The resolving power of the spectra was R$\sim$50000
and the exposure times were chosen to achieve a S/N ratio larger than
200. The reduction and calibration of the spectra were carried out with the
standard CRIRES pipeline procedures. Hot standard stars at similar air
mass were observed immediately after each target object to properly
remove telluric lines using the task {\it telluric} within the IRAF
software package.

Fluorine abundances were derived from an unblended line R9 (1-0) of
the HF molecule at $\lambda\sim 2.3358~\mu$m (\cite{Abia2009}).  We used the classical
method of spectral synthesis in LTE for the abundance analysis.
Theoretical spectra were computed with the TURBOSPECTRUM code
(\cite{alv}, and further improvements) and convolved with Gaussian
functions to mimic the corresponding instrumental profiles.
To this purpose, we used model atmospheres computed from the new
version of the MARCS code (\cite{Gustafsson08}). The atomic and
molecular line lists are the same used previously by
\cite{Abia2009}. They have been calibrated using the spectra of the Sun and Arcturus
(the adopted solar abundance of F is 4.56~dex according to \cite{Asplund}).
Accurate stellar parameters and iron abundances for all the program
stars, presented in Table~1,  were taken from \cite{Santos04}, \cite{Santos05} and
\cite{Sousa06}.  These parameters were derived from a homogeneous
analysis of optical spectra, thus minimizing the relative errors.

Figure~1 shows the comparison between the observed spectrum of one of our targets
(HD131977) and our best fit synthetic spectrum in the region of the HF R9 line.
In addition, two synthetic spectra, corresponding to the errors of the fit for the F abundance
($\pm$ 0.1 dex with respect to the best fit spectrum) are plotted in a zoom window of the same figure.

In addition to the CRIRES data, and for six of our target stars, we
used FEROS spectra from the ESO archive to derive the abundances of
two light s-elements: yttrium and zirconium. To this purpose, two
wavelength regions, 6120 to 6150 \AA\ and 6420 to 6450 \AA, were
investigated. Within these regions spectral features for iron, yttrium
and zirconium were analysed using the spectral synthesis programme MOOG
(\cite{MOOG}) with MARCS stellar atmosphere models
(\cite{Gustafsson08}).  We tried to measure other s-element abundances,
as Ba and Nd, but the quality of the lines was not good enough for the
analysis. The Eu~II line at 6437 \AA\ was not detected, as expected for a
solar Eu abundance at this effective temperature range. We therefore
suppect that our target stars are probably not enriched in r-elements, which are
produced in Type~II supernovae, with an upper limit of 0.3~dex in 
[Eu/H]. 
The signal-to-noise ratio (S/N) of the FEROS spectra was
between 65 and 85 per pixel, depending on the star.
The atomic and molecular linelists were
collated using the most recent laboratory values where possible, as
outlined in \cite{Worley2010} and references therein. For the
purposes of this analysis, the linelists were calibrated to the Sun
using the MARCS solar atmosphere model. The Fe abundances determined
by spectrum synthesis, from both Fe I and Fe II lines, were in good agreement with the metallicities
determined previously for these stars by \cite{Santos04}, \cite{Santos05}, and
\cite{Sousa06}.

\begin{figure*}[h]
\includegraphics[width=13cm,height=13cm]{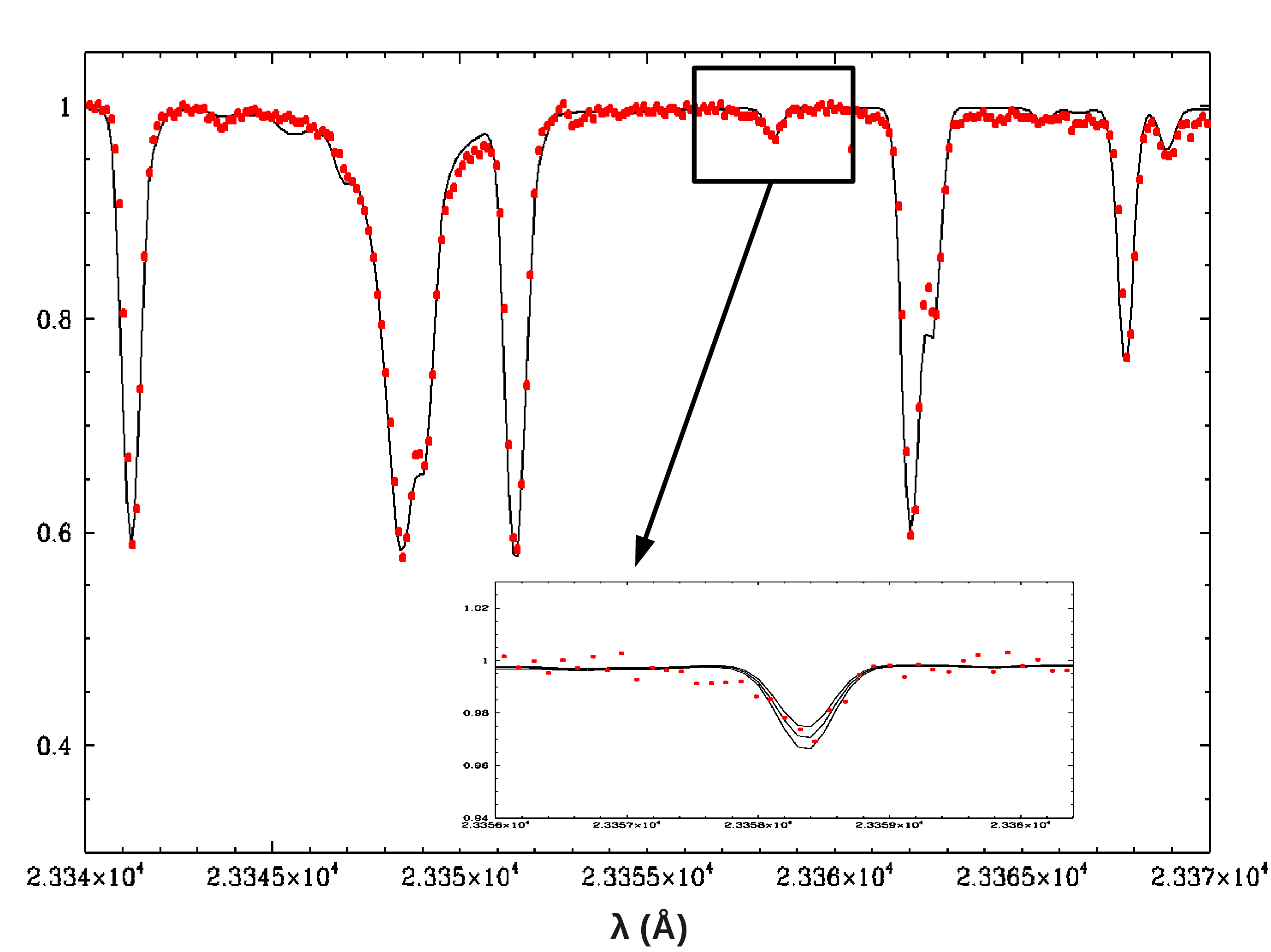}
\caption{CRIRES spectrum of HD131977 (red points) and our best fit synthetic spectrum (black line). The embedded window shows a zoom of the
line R9 (1-0) of the HF molecule  at $\lambda\sim  2.3358~\mu$m. The central synthetic spectrum corresponds to the fluorine abundance of our best fit, while the other two lines show the flux variation for an error of 0.10 dex with respect to the best fit.}
\label{}
\end{figure*}

%

\section{Results on fluorine and s-element abundances}

The derived abundances of fluorine, yttrium, and zirconium for the target stars are presented in Table 1.
For each element, the abundance uncertainty was calculated taken into account the error in the fit
(as shown in Fig. 1) and the individual sensitivities of the derived abundances to the adopted
effective temperature, surface gravity, metallicity, and microturbulence velocity within the corresponding
uncertainties (see Table~1). The resulting abundance uncertainties for these
different sources of error were then summed in quadrature to determine the final formal
uncertainty that is equal to 0.20~dex for [F/H] and 0.16~dex for [Y/Fe] and [Zr/Fe].

\begin{table*}[h]
\caption{Adopted atmospheric parameters (from \cite{Santos04}, \cite{Santos05}, and
\cite{Sousa06}) and derived fluorine, yttrium, and zirconium abundances for the target stars.}
\label{}
\centering
\begin{tabular}{l c c c c c c c}
\hline\hline
Target ID & Teff (K) & logg (dex) &  [Fe/H] (dex) & $ \xi $ (km/s) & log$\epsilon$(F) (dex) & [Y/Fe] (dex) & [Zr/Fe] (dex) \\ 
\hline
HD50281   & 4658 $\pm$ 56   & 4.32 $\pm$ 0.24   & -0.04 $\pm$ 0.07   & 0.64 $\pm$ 0.15   & 4.53 $\pm$ 0.20 & 0.35 $\pm$ 0.17 & 0.36 $\pm$ {\bf 0.16}\\ 
HD65486   & 4660 $\pm$ 66   & 4.55 $\pm$ 0.21   & -0.33 $\pm$ 0.07   & 0.82 $\pm$ 0.16   & 4.47 $\pm$ 0.20  & --         & --         \\ 
HD85512   & 4505 $\pm$ 176  & 4.71 $\pm$ 0.96   & -0.18 $\pm$ 0.19   & 0.32 $\pm$ 1.15   & 4.73 $\pm$ 0.20  & --         & --         \\ 
HD101581  & 4646 $\pm$ 96   & 4.80 $\pm$ 0.39   & -0.37 $\pm$ 0.09   & 0.58 $\pm$ 0.35   & 4.61 $\pm$ 0.20  & 0.37 $\pm$ 0.17 & 0.32 $\pm$ {\bf 0.16}\\ 
HD111261  & 4529 $\pm$ 62   & 4.44 $\pm$ 0.64   & -0.35 $\pm$ 0.08   & 0.78 $\pm$ 0.17   & 4.44 $\pm$ 0.20  & --         & --         \\ 
HD131977  & 4693 $\pm$ 80   & 4.36 $\pm$ 0.25   &  0.07 $\pm$ 0.10   & 0.97 $\pm$ 0.16   & 5.16 $\pm$ 0.20  & 0.33 $\pm$ 0.17 & 0.17 $\pm$ {\bf 0.16}\\ 
HD156026  & 4568 $\pm$ 94   & 4.67 $\pm$ 0.76   & -0.18 $\pm$ 0.09   & 0.60 $\pm$ 0.26   & 4.41 $\pm$ 0.20  & 0.33 $\pm$ 0.17 & 0.20 $\pm$ {\bf 0.16}\\ 
HD209100  & 4629 $\pm$ 77   & 4.36 $\pm$ 0.19   & -0.06 $\pm$ 0.08   & 0.42 $\pm$ 0.25   & 4.75 $\pm$ 0.20  & 0.30 $\pm$ 0.17 & 0.27 $\pm$ {\bf 0.16}\\ 
HD216803  & 4555 $\pm$ 87   & 4.53 $\pm$ 0.26   & -0.01 $\pm$ 0.09   & 0.66 $\pm$ 0.28   & 4.64 $\pm$ 0.20  & 0.15 $\pm$ 0.17 & 0.06 $\pm$ {\bf 0.16}\\ 
\end{tabular}
\end{table*}

Figure 2 shows our derived fluorine abundances with respect to iron as a function of the iron content
taken from \cite{Santos04}, \cite{Santos05}, and \cite{Sousa06}. The corresponding value for the Sun is
indicated as a reference. Taken into account the error bars, at least three of the 
analysed stars seem to be fluorine-enhanced with respect to the Sun (up to 0.35-0.55~dex). No correlation with metallicity
can be detected: the two more fluorine-overabundant stars of the sample are HD131977 with 
[Fe/H]$=$+0.07$\pm$0.10, and HD101581 with [Fe/H]$=$-0.37$\pm$0.09. The corresponding Pearson correlation coefficient
between [F/Fe] and [Fe/H] is also very low (r=-0.22).

\begin{figure*}[ht]
\centering
\includegraphics[width=12cm,height=10cm]{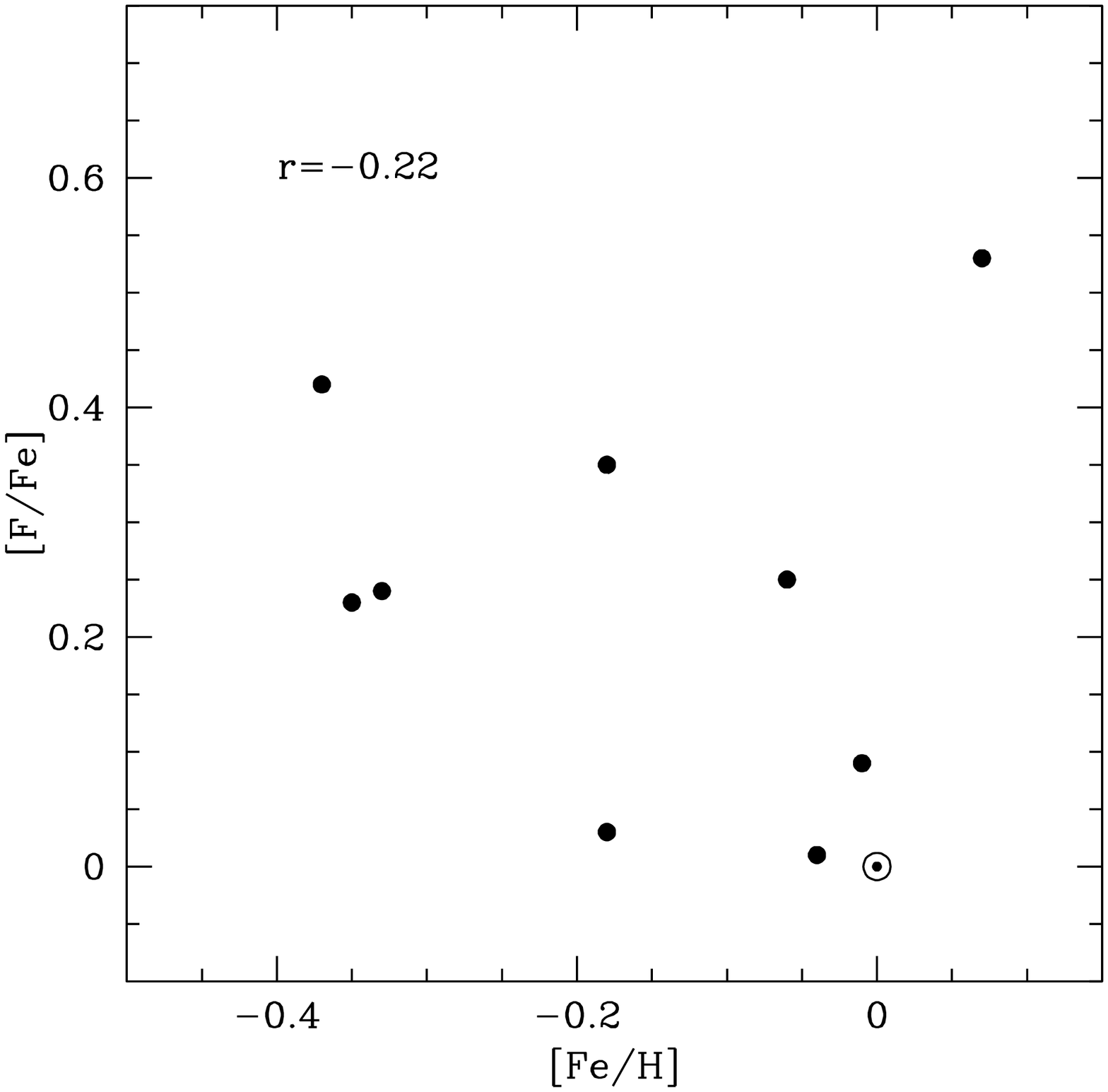}
\caption{Derived fluorine abundances with respect to iron (mean error of 0.29~dex) as a function of the iron content (black dots).
The corresponding value for the Sun is shown as a reference. The lack of correlation between the [F/Fe] content and the
[Fe/H] abundance is indicated by the low value of the Pearson correlation coefficient {\it r}.}
\label{}
\end{figure*}

The derived yttrium and zirconium abundances have allowed us to check the existence of a correlation between the fluorine
production and the s-element production, and therefore, to test whether the fluorine present in the material
from which our target stars were formed comes mainly from AGB stars. As shown in Table 1, the determined values of
[Y/Fe] and [Zr/Fe] for each star agree within the errors. A mean s-element abundance
was therefore calculated from them and compared to the [F/Fe] abundance (Fig. 3 left panel). We conservatively assume an
error of 0.10~dex for the mean s-element abundance with respect to iron. The fluorine-enriched stars clearly also 
seem to be the most s-element enriched (with $<[s/Fe]>$ equal to 0.27$\pm$0.10~dex and 0.36$\pm$0.10~dex for the two
more F-enriched stars), suggesting a common nucleosynthesis origin. Even if this relation
is not so straightfoward for two of the stars (which seem to be s-element overabundant but have very small [F/Fe] 
enhancements) 
a general correlation cannot be excluded within the errors. The dispersion in our measurements,
coming mostly from the F abundance, could explain the rather low correlation coefficient value of r=0.48.

However, to test the possible Type~II supernovae origin of fluorine, it is interesting to compare
our F abundances with the alpha-element content, as the latter is a tracer of Type~II supernovae nucleosynthesis.
Homogeneous abundances for several alpha-elements and iron-peak elements were available from
Gilli et al. (2006) for all the targets of our sample. In addition, our analysis and that of \cite{Gilli06} 
rely on the same atmospheric parameter values for the stars, taken from \cite{Santos04}, \cite{Santos05}, and \cite{Sousa06}.
This parameter consistency allows us to minimise any systematic errors in the element abundance comparisons.
The right panel of Figure~3 shows our F abundances plotted as a function of the \cite{Gilli06} [$\alpha$/Fe] content. 
The correlation between the [F/Fe] abundance and the [$\alpha/Fe$] is again rather low (r=0.54), and
in contrast to the s-element abundances, the F-enhanced stars do not seem to be alpha-element enriched with
respect to iron (the dispersion in the [$\alpha/Fe$] values is very low and the [$\alpha$/Fe] content for the two
more F enhanced stars is equal to 0.03$\pm$0.10~dex and 0.08$\pm$0.10~dex). This result suggest that 
Type~II supernovae should be excluded as the main F contributors for these stars.

\begin{figure*}[h]
\centering
\includegraphics[width=12cm,height=10cm]{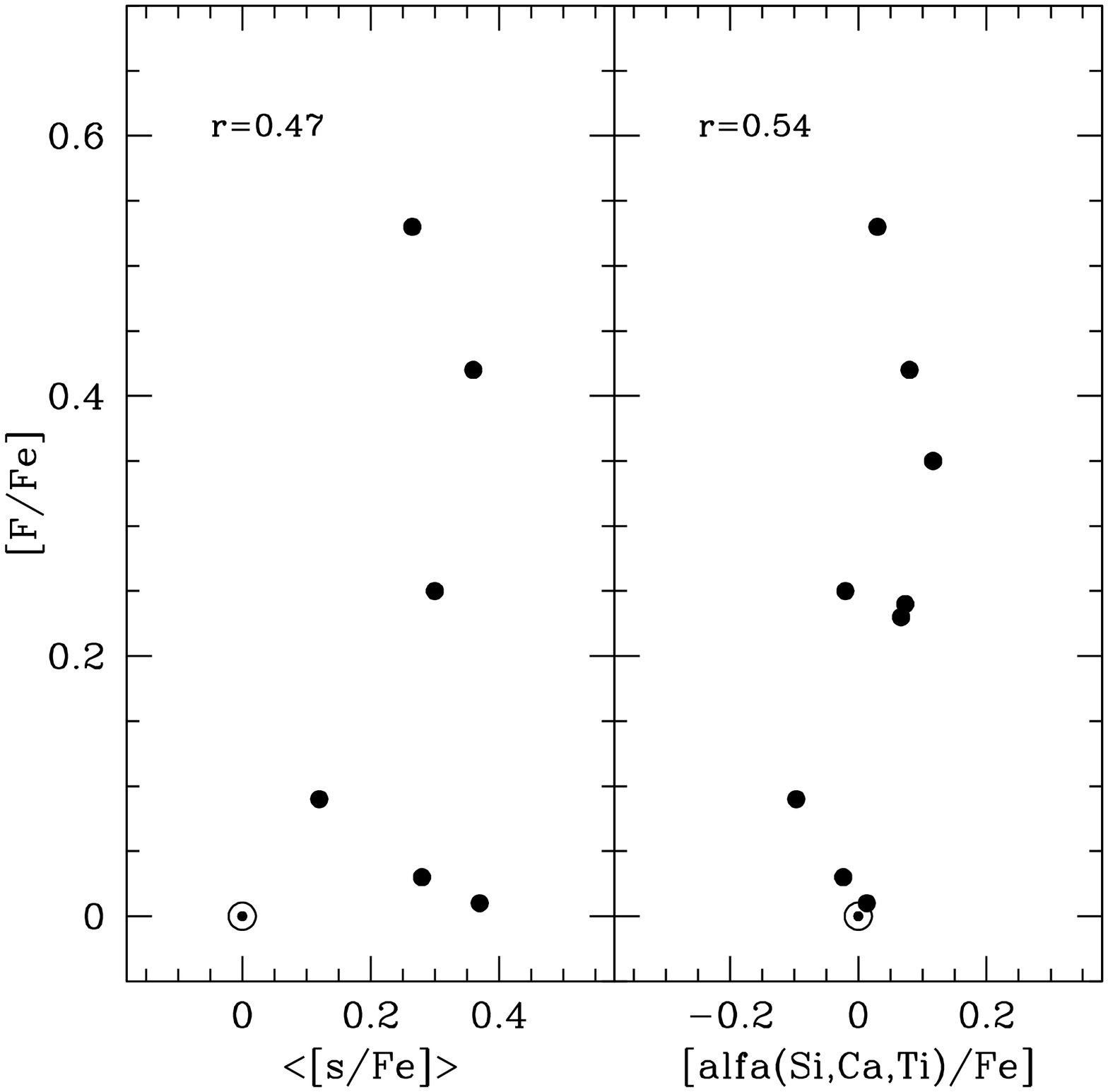}
\caption{Measured fluorine abundances as a function of the mean s-element content (derived from yttrium and zirconium; left panel) 
and the mean [$\alpha/Fe$] (derived from Si, Ca, and Ti measurements by Gilli et al. 2006; right panel). The mean error in [F/Fe]
is $\sim$ 0.29~dex, and 0.1~dex for $<[s/Fe]>$ and [$\alpha$/Fe]. As for Fig. 2, the solar value and the Pearson correlation
coefficient are indicated. 
}
\label{}
\end{figure*}

Finally, we searched for possible correlations of the fluorine abundance with several refractory element ones,
to determine whether there is no link between the F enhancements and the stellar metallicity.
Figure 4 shows our determined F abundance as a function of the Sc, V, Cr, Mn, Co, and Ni content. The corresponding
Pearson correlation co-efficient between the F abundance and each element abundance is also indicated. The
solar values are again marked as a reference. No significant correlations seem to exist, except maybe 
between F and Co, but this relation is driven particularly by the values found for the star HD131977 and
is clearly not statistically robust.


\begin{figure*}[h]
\centering
\includegraphics[width=17cm,height=17cm]{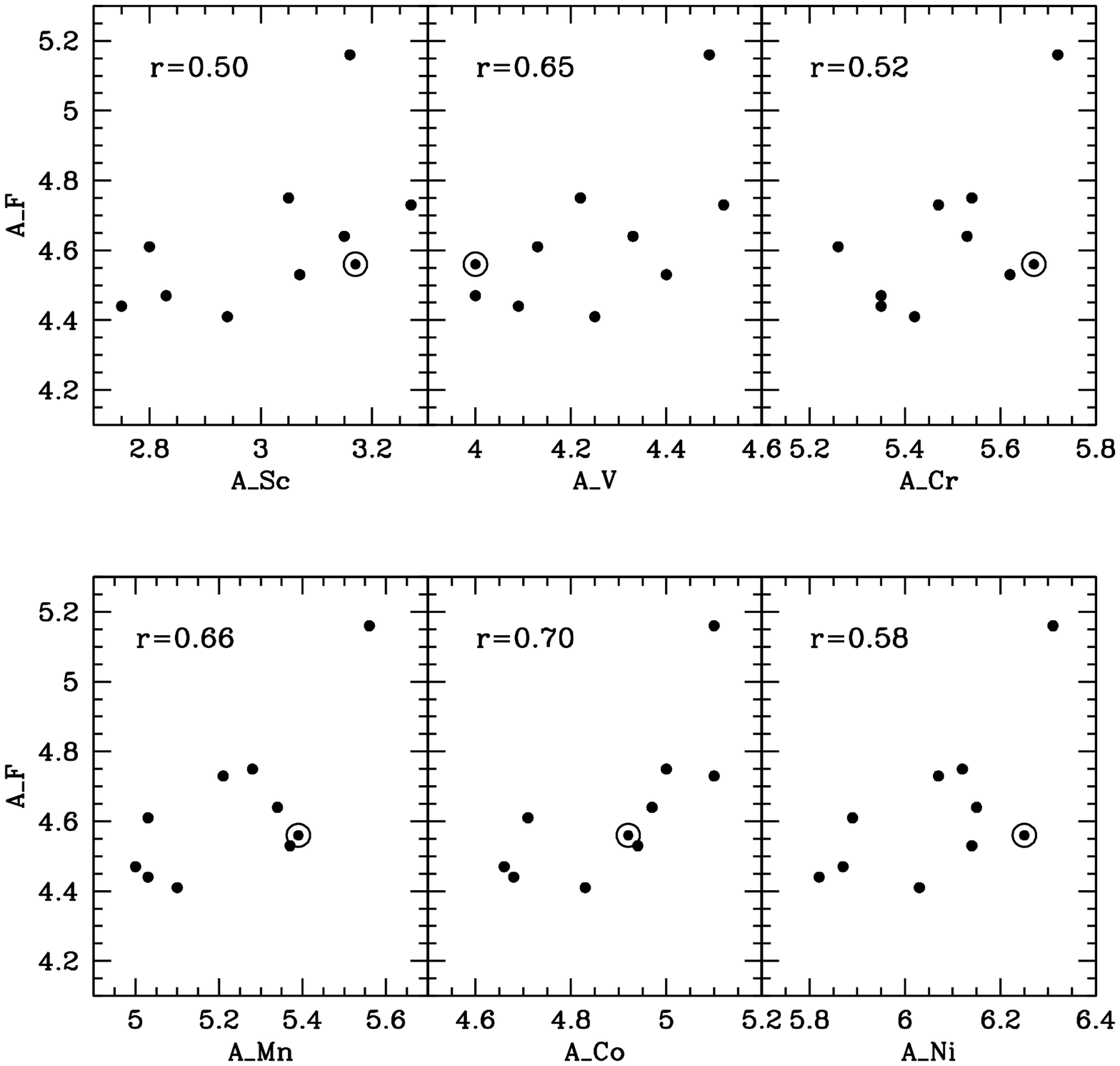}
\caption{Derived fluorine abundances as a function of several refractory elements abundances. Simbols are those of Fig. 2.
The corresponding
Pearson correlation coefficient {\it r} between the F abundance and the related element one is also indicated.}
\label{}
\end{figure*}


\section{Conclusions}
We have derived, for the first time, fluorine abundances for a set of nine cool main-sequence dwarfs of the solar
neighbourhood, from an unblended line of the HF molecule at 2.3 microns. The studied
stars have the advantange of being exempt of external pollution and internal mixing processes
that can affect the observed F abundance patterns.
Several of the analysed stars seem to be fluorine enhanced with respect to the Sun, but
no correlation with the stellar iron content is found.
We took advantage of FEROS archive spectra, available for six of the targets, to 
determine their yttrium and zirconium abundances and to test whether AGB fluorine nucleosynthesis 
can account for the observed F enhancements. The fluorine
enriched stars clearly also seem to be s-element enriched, suggesting that AGB stars could
be the origin of the F present in the studied dwarfs of the solar neighbourhood.
Nevertheless, the correlation between [F/Fe] and the s-elements is rather low and possibly blurred by 
the uncertainties in the F abundance measurements.
Finally, our derived F abundances were compared to the alpha-element and iron-peak
element abundances of \cite{Gilli06}, which were determined using the same
set of stellar atmospheric parameters. Type II core-collapse supernovae are unlikely to be
the dominant producers of F for our targets, as no correlation 
seems to exist between [F/Fe] abundance and the [$\alpha/Fe$] ratio. This is consistent
with the fact that our targets not being r-element enriched (upper limits to Eu of 0.3~dex
have been found). In addition, the 
absence of a clear correlation between the abundances of F and the iron-peak elements
confirms that the observed F enhancements are not related to the stellar metallicity,
in the explored high metallicity range.
In summary, the observed fluorine, s-element and alpha-element abundance patterns
suggest that AGB stars are the dominant source of fluorine in the studied dwarf 
stars of the solar neighbourhood.

\begin{acknowledgements}
We are grateful to Carlos Abia for his useful comments about the paper.
C. Worley acknowledges the support of the French Centre National d'\'Etudes Spaciales (CNES).

\end{acknowledgements}

\end{document}